\documentclass[a4paper]{article}

\usepackage{INTERSPEECH2020}
\usepackage{algorithm}
\usepackage{algorithmic}
\usepackage{multirow}

\title{Inaudible Adversarial Perturbations for Targeted Attack in Speaker Recognition}
\name{Qing Wang, Pengcheng Guo, Lei Xie}
\address{
  Audio, Speech and Language Processing Group (ASLP@NPU), \\
  School of Computer Science, Northwestern Polytechnical University, Xi'an, China}
\email{\{qingwang, pcguo, lxie\}@nwpu-aslp.org}

\begin{document}

\maketitle
\begin{abstract}

  Speaker recognition is a popular topic in biometric authentication and many deep learning approaches have achieved extraordinary performances. However, it has been shown in both image and speech applications that deep neural networks are vulnerable to adversarial examples. In this study, we aim to exploit this weakness to perform targeted adversarial attacks against the x-vector based speaker recognition system. We propose to generate inaudible adversarial perturbations achieving targeted white-box attacks to speaker recognition system based on the psychoacoustic principle of frequency masking. Specifically, we constrict the perturbation under the masking threshold of original audio, instead of using a common $l_p$ norm to measure the perturbations. Experiments on Aishell-1 corpus show that our approach yields up to 98.5\% attack success rate to arbitrary gender speaker targets, while retaining indistinguishable attribute to listeners. Furthermore, we also achieve an effective speaker attack when applying the proposed approach to a completely irrelevant waveform, such as music.

\end{abstract}

\section{Introduction}
In recent years, attacks and defenses of speaker recognition systems have attracted more and more attention. As one of the most prominent biometric authentication methods, the security of speaker identification system is extremely important. 
Prior works have found speaker recognition systems are not only facing the spoofing attacks~\cite{wu2015asvspoof, wu2015spoofing, wu2016anti} including impersonation, replay, speech synthesis, as well as voice conversion, while adversarial attacks are also be able to affect speaker recognition systems. In~\cite{das2020attacker}, Das \textit{et al.} gave an overview of the attacker's perspective on speaker verification.

Adversarial attacks are usually conducted by adversarial examples, which are designed by constructing imperceptible perturbations to lead a mis-classification. Adversarial examples were first proposed by Szegedy \textit{et al.}~\cite{szegedy2013intriguing} in computer vision tasks, which show that a certain network is vulnerable to a crafted small perturbation in the training set. Goodfellow \textit{et al.}~\cite{goodfellow2014explaining} proposed an effective approach, fast gradient-sign method (FGSM), to generate adversarial examples through the linearization of the loss function. Since then, various experimental results have shown that adversarial examples can successfully influence a variety of models~\cite{kurakin2016adversarial,miyato2018virtual}.

Apart from the applications in image tasks, speech-related tasks could also be affected by adversarial examples. 
There has been plenty of work focused on attacking automatic speech recognition (ASR) systems using adversarial examples. In~\cite{carlini2018audio}, Carlini \textit{et al.} demonstrated the effectiveness of targeted audio adversarial examples on a end-to-end ASR system. With optimization-based attacks, they were able to turn any audio waveform into any target transcription. Instead of using a $l_p$ norm to measure the maximum perturbation introduced as above, Sch{\"o}nherr \textit{et al.}~\cite{schonherr2018adversarial} introduced a new type of adversarial examples based on psychoacoustic hiding and attacked the Kaldi ASR system~\cite{povey2011kaldi} successfully. Next, Qin \textit{et al.}~\cite{qin2019imperceptible} extended this idea and developed effectively imperceptible audio adversarial examples by leveraging the psychoacoustic principle of auditory masking.

In speaker recognition area, adversarial examples could also be used to attack and to defend the system. In~\cite{kreuk2018fooling}, Kreuk \textit{et al.} used adversarial examples for fooling a speaker verification (SV) system by adding a peculiar noise to the original speaker examples. In our previous work~\cite{wang2019adversarial}, we added adversarial perturbations on feature-level to conduct a non-targeted attack to SV system. We also explored using adversarial examples for model regularization and improved the robustness of the SV system. Xie \textit{et al.}~\cite{xie2020real} made the DNN based speaker recognition system can identify the speaker as any target label by adding audio-agnostic universal perturbations on speakers' voice input. In~\cite{li2020universal}, Li \textit{et al.} proposed to generate universal adversarial perturbations (UAPs) by learning the mapping from the low-dimensional normal distribution to the universal perturbation subspace via a generative model. However, the aforementioned adversarial examples are mostly restricted to make a slight change of original signal in audio sampling points, without considering the human perceptibility of sound.

In this study, we were inspired by the work in~\cite{schonherr2018adversarial,qin2019imperceptible} and propose to generate inaudible adversarial perturbations for targeted attacking speaker recognition directly on wave-level. We use the structure of the x-vector speaker recognition system proposed in~\cite{snyder2018x} as our baseline to conduct targeted white-box attacks. To generate the inaudible adversarial perturbations, we adopt the frequency masking concept where one faint but audible sound becomes inaudible in the presence of another louder audible sound.
Our experimental results based on Aishell-1~\cite{bu2017aishell} corpus demonstrate that the inaudible adversarial perturbations can achieve better targeted attack performance than previous $l_p$ norm based adversarial examples. To further compare the frequency masking based approach with previous ones, we also evaluate them from both subjective and objective metrics. Results show that the adversarial perturbations generated by proposed methods are more inaudible, even with larger absolute energy.
Finally, we attempt to conduct targeted attacks using the music portion of the MUSAN  corpus~\cite{snyder2015musan}, which is a completely irrelevant non-speech dataset. Experiments show that even non-speech can also achieve a high speaker attack success rate.
\begin{figure*}[t]
  \centering
  \includegraphics[width=15cm]{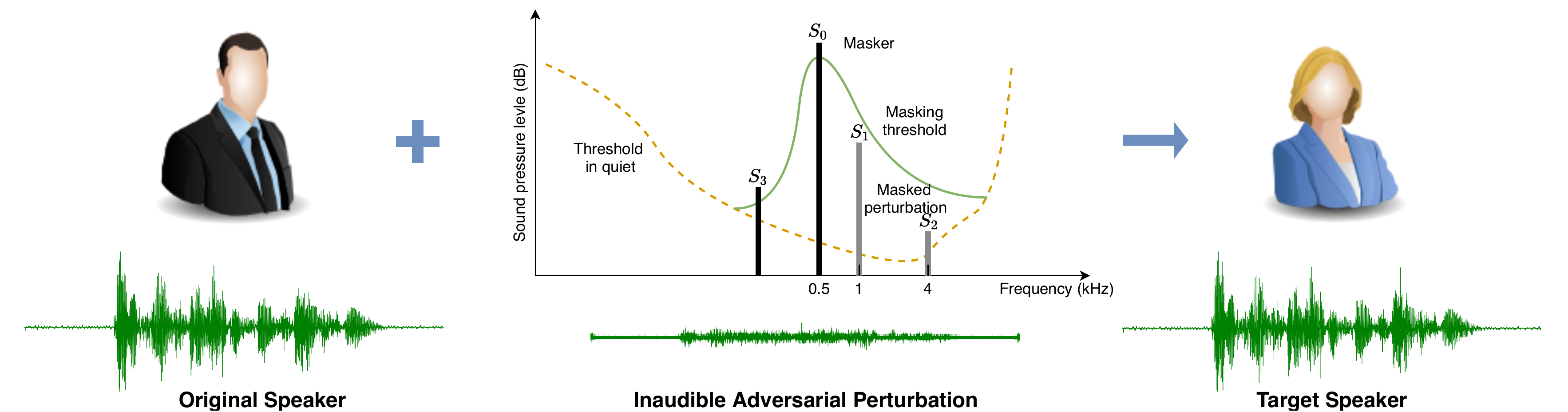}
  \caption{An overview of the generation of adversarial examples based on frequency masking. }
  \label{fig:overview}
\end{figure*}

The rest of the paper is organized as follows. In Section 2, we detail the generation of the inaudible adversarial perturbations. In Section 3, we describe the experimental setup. Experimental results and analysis are presented in Section 4. We conclude in Section 5.

\section{Inaudible adversarial perturbations}
In this section, we introduce how we generate the inaudible perturbations that can conduct targeted speaker attacks. Figure~\ref{fig:overview} shows an overview of the generation of adversarial examples base on frequency masking.

\subsection{Adversarial example generation}
An adversarial example is defined as an instance with imperceptible, intentional perturbation that causes a well-trained model to make a false prediction. Conventional approaches to generate adversarial perturbations are typically by performing gradient descent w.r.t the input sample. Specifically, given an input speech $x$, its label speaker $y$, an arbitrary target label $y'$ and a well-trained speaker recognition model $f(\cdot)$, the adversarial perturbation $\delta$ can be generated by

\begin{equation}\label{commo_method}
    \begin{aligned}
        & \min L_{CE}(f(x+\delta), y'), \\
        & s.t. \quad \lVert \delta \rVert < \epsilon,        
    \end{aligned}
\end{equation}
where $y' \neq y$ and $L(\cdot)$ is the loss function. The hyperparameter $\epsilon$ is used to control the maximum perturbation generated.

\subsection{Frequency masking}
Our goal is to generate indistinguishable adversarial perturbations in the human perceptibility of audio, instead of maintaining a slight noise to the clean speech sample points. In order to achieve that, we utilize the idea of \textit{frequency masking}, which refers to the phenomenon that one faint but audible sound (the maskee) becomes inaudible in the presence of another louder audible sound (the masker)~\cite{lin2015principles}. Therefore, we can modify adversarial perturbations to be inaudible, as long as the perturbation falls under the masking threshold of the original speech. In~\cite{lin2015principles}, Lin \textit{et al.} investigated the algorithm of computing masking threshold, which consists of 3 steps. 

\vspace{5pt}
\noindent \textbf{STEP 1: Identifications of maskers}
\vspace{5pt}

In order to obtain the frequency masking threshold of the original speech, raw audio signals from the time domain are first converted into time-frequency representations by short-time Fourier transform (STFT). The output of STFT $s_x(k)$ refers the $k$-th bin of the spectrum at frame $x$. Then, the power spectral density (PSD) of $s_x(k)$ can be computed as
\begin{equation}
    P_x(k) / {\rm dB} = 10\log_{10} \vert \frac{1}{N}s_x(k) \vert^2.
\end{equation}
After that, the PSD estimate $P_x(k)$ is normalized to a sound pressure level (SPL) of 96 dB,
\begin{equation}
    \overline{P}_x(k) / {\rm dB} = 96 - max\{P_x(k)\} + P_x(k).
\end{equation}

The normalized PSD estimate of reasonable maskers must satisfy three constraints. First is local maxima,
\begin{equation}
    \overline{P}_x(k) \geq \overline{P}_x(k) \quad {\rm and} \quad \overline{P}_x(k) \geq \overline{P}_x(k).
\end{equation}
Secondly, they should be larger than the absolute threshold of hearing (ATH),
\begin{equation}
    \overline{P}_x(k) \geq {\rm ATH(k)}.
\end{equation}
Finally, any group of maskers should keep a maximum amplitude within 0.5 Bark (a psychoacoustically-motivated frequency scale) and only the masker with the highest SPL is retained,
\begin{equation}
    \overline{P}_{x_1, x_2}(k) = \mathop{\arg\max}_{k_0 \in [-0.5, 0.5]} \overline{P}_{x_1, x_2}(k + k_0).
\end{equation}{}

Since the masking effect is additive in the logarithmic domain, the SPL of each masker can be further smoothed by
\begin{equation}
    \overline{P}_x(\overline{k}) = 10\log_{10}[10^{\frac{\overline{P}_x(k-1)}{10}} + 10^{\frac{\overline{P}_x(k)}{10}} + 10^{\frac{\overline{P}_x(k+1)}{10}}].
\end{equation}{}
\vspace{5pt}
\noindent \textbf{STEP 2: Calculation of individual masking thresholds}
\vspace{5pt}

An individual masking threshold $T[b(j), b(i)]$ means that the masker at frequency index $j$ contributes to the masking effect on the maskee at frequency index $i$, where $b(j)$ and $b(i)$ are the masker and maskee’s frequencies in Bark scale. The individual masking thresholds can be calculated as:
\begin{equation}
    T[b(j), b(i)] / {\rm dB} = \overline{P}_x[b(j)] + \Delta[b(j)] + {\rm SF}[b(j), b(i)],
\end{equation}
where $\Delta[b(j)] = -6.025-0.275b(j)$ and ${\rm SF}[b(j), b(i)]$ is a two-slop spread function.

\vspace{5pt}
\noindent \textbf{STEP 3: Calculation of global masking threshold}
\vspace{5pt}

After the individual masking thresholds are obtained, the global masking threshold can be calculated by combining them with the absolute threshold of hearing. The global masking threshold at frequency index $i$ is calculated according to
\begin{equation}
    T_G(i) / {\rm dB} = 10\log_{10}[10^{\frac{{\rm ATH(i)}}{10}} + \sum_{j=1}^{N_M}10^{\frac{[b(j), b(i)]}{10}}],
\end{equation}
where ${\rm ATH(i)}$ is the SPL of threshold in quiet at frequency index $i$, $N_M$ is the number of maskers, and $T[b(j), b(i)]$ is corresponding individual masking threshold. Readers can get more detail about the calculation of masking threshold in~\cite{lin2015principles}.

\subsection{Optimization procedure}

Given an input speech $x$, its label speaker $y$, an arbitrary target speaker label $y'$, where $y \neq y'$, and a well-trained x-vector speaker recognition model $f(\cdot)$, the additional loss function to modify the perturbation fall under the masking threshold can be defined as
\begin{equation}\label{th_loss}
    L_{TH}(x, \delta) = \mathbb{E}_{k} \max \{\overline{P}_{\delta}(k) - T_G(k), 0\},
\end{equation}
where $\overline{P}_{\delta}(k)$ means the normalized PSD estimated of $\delta$ at the $k$-th frequency bin. The inaudible adversarial perturbation $\delta$ can be generated by
\begin{equation}\label{tot_loss}
    \min L(x, \delta, y') = L_{CE}(f(x+\delta), y') + \alpha \cdot L_{TH}(x, \delta),
\end{equation}
where $L_{CE}$ aims to make the adversarial examples fool the well-trained speaker recognition system into predicting an arbitrary target label and the $L_{TH}$ constrains the normalized PSD estimate of perturbation to be inaudible. The $\alpha$ is a hyper-parameters to scale different losses.

The whole optimization procedure is separated into two stages. In \textbf{Attack Stage1}, we focus on finding a relative small perturbation using a common $l_p$ norm based algorithm as defined in Eq.~\eqref{commo_method}. The $\delta$ is initialized to a zero vector and $\epsilon$ is gradually reduced from a large value. For each iteration, $\delta$ is updated by
\begin{equation}
    \delta \leftarrow clip_{\epsilon}(\delta - lr_1 \cdot sign(\nabla_{\delta}L_{CE}(f(x+\delta), y'))).
\end{equation}
In \textbf{Attack Stage2}, we further optimize above perturbation by introducing frequency masking based loss as defined in Eq.~\eqref{tot_loss}. The $\alpha$ starts from $0.05$ and adaptively updated based on the performance of attack. For each iteration, $\delta$ is updated to be inaudible through:
\begin{equation}
    \delta \leftarrow \delta - lr_2 \cdot \nabla_{\delta}L(x, \delta, y').
\end{equation}{}

\section{Experimental setup}
\subsection{Dataset}
We use the Mandarin Aishell-1 corpus~\cite{bu2017aishell} as the evaluation data set. The entire corpus contains 400 speakers (214 female, 186 male), sampled at 16kHz, including training, development and test sets, without speaker overlapping. 
Training set is used in x-vector baseline training, while test set is used to evaluate the baseline system. 
For conducting inaudible adversarial targeted attacks, we randomly choose 10 female speakers (denoted as F) and 10 male speakers (denoted as M) from the training set, each with 100 utterances, as the original speaker set. Another 10 female speakers (denoted as F') and 10 male speakers (denoted as M') are selected as the attack targets. We assign these selected sets into 4 test modes.
The first one is using 10 male original speakers to attack 10 male target speakers, denoted as M2M'. Similarly, the other three test modes are M2F', F2M' and F2F'. 

Besides, we use the music portion of MUSAN~\cite{snyder2015musan} corpus as our non-speech dataset, which consists of western art music (e.g., Baroque, Romantic, and Classical) and popular genres (e.g., jazz, bluegrass, hip-hop, etc). We randomly choose 200 pieces of western art music and cut them into 1000 pieces of 6 seconds short segments. This subset is used as the original wave to attack the selected male target speakers.

\subsection{Experimental setup}
\subsubsection{Baseline}
We use x-vector system~\cite{snyder2018x} as our baseline. The 30-dimensional Mel-frequency cepstral coefficients (MFCC) features are extracted as the input for all experiments. 
The configuration of x-vector network is exactly the same as in~\cite{snyder2018x}: a 5-layer TDNN with ReLU followed by batch normalization is used for extracting frame-level hidden features. The number of hidden nodes is 512 and the dimension of frame-level hidden features for pooling is 1500. Each frame-level feature is generated from a 15-frame context of acoustic features. Pooling layer aggregates frame-level features, followed by 2 fully-connected layers with ReLU activation functions, batch normalization, and a Softmax output layer. The EER of the x-vector baseline system is 4.27\%. Note that we use the whole sentence as input instead of using chunks as in~\cite{snyder2018x}, because we need to compute the gradient w.r.t the sentence-level perturbation.

After training the x-vector baseline system, we calculate the speaker prediction for the original utterances with their true labels. The accuracy for the M set is 95.9\%, while the accuracy for the F set is 97.9\%. We also calculate the prediction accuracy for the original utterances with assigned target speakers. All the results of the four test modes are 0.00\%.

\subsubsection{Inaudible adversarial perturbations}
We first compute the STFT of original speech to get the time-frequency representations. The window type of STFT is the modified Hann window with a length of 2048 and a hop length of 512. In Attack Stage1, the learning rate $lr_1$ is set to be $100$ and the $\delta$ will be updated $3000$ times for each mini-batch. We use the $l_{\infty}$ norm to measure the perturbation bound. The $\epsilon$ starts from 2000 and will multiply $0.8$ when attacking successfully. In Attack Stage2, the learning rate $lr_2$ is $1$ and the total training step for each mini-batch is $1000$. The scale parameter $\alpha$ begins with $0.05$ and will increase to $1.2\alpha$ when attacking successfully or decrease to $0.8\alpha$ when fails. All systems are implemented using PyTorch~\cite{paszke2017automatic} and optimized by Adam optimizer~\cite{kingma2014adam}.

\subsubsection{Evaluation metrics}
We use various metrics to measure the performance of proposed method. First, we compute the attack success rate to evaluate the performance of targeted attacks in speaker recognition. Formally, the accuracy is computed as:
\begin{align}
    Acc = N_s/N,
\end{align}
where $N$ is the total number of utterances we used to test and $N_s$ refers to the number of utterances attacking. Besides, perceptual evaluation of speech quality (PESQ)~\cite{rix2001perceptual} and signal-to-noise ratio (SNR) are also computed to measure the distortion of generated adversarial examples. Finally, we also conduct a subjective evaluation to evaluate the adversarial examples from the human perceptibility of audio. successfully.

\section{Experimental results and analysis}

\subsection{Inaudible adversarial targeted attack}

In Table~\ref{tab:1}, we calculate the attack success rate for all the four test modes.
As we separated the optimization procedure into two stages in Section 2.3. We will test the adversarial examples generated in these two stages, donated as Attack Stage1 and Attack Stage2, respectively.
System Attack Stage1 is we conduct attack using the adversarial examples generated in Attack Stage1, which just focus on finding a small perturbation. And the targeted attack successfully affected the speaker model in 72.6\%, 73.8\%, 73.3\% and 71.3\%  of cases in these four test modes.
For System Attack Stage2, the frequency masking method is used in generating inaudible adversarial perturbations.
The rates of successful targeted attacks in four test modes are 98.5\%, 97.6\%, 96.7\% and 93.8\%.
In this experiment, adversarial examples from both attack stages can successfully conduct targeted speaker attacks. We can achieve a higher attack success rate in System Attack Stage2, which indicates the effectiveness of the inaudible adversarial perturbations in targeted attacks.
\begin{table}[!ht]

  \caption{Attack success rate (\%) of Attack Stage1 and Attack Stage2 on each test mode.}

  \label{tab:1}
  \centering

  \begin{tabular}{c c c c c }

    \toprule
    \multicolumn{1}{c}{\textbf{System}} &
    \multicolumn{1}{c}{\textbf{M2M'}} &
    \multicolumn{1}{c}{\textbf{M2F'}} &
    \multicolumn{1}{c}{\textbf{F2M'}} &
    \multicolumn{1}{c}{\textbf{F2F'}} \\
    \midrule
    \textit{Attack Stage1} & 72.6 & 73.8 & 73.3 & 71.3   \\
    \textit{Attack Stage2} & \textbf{98.5} & \textbf{97.6} & \textbf{96.7} & \textbf{93.8}   \\
    \bottomrule

  \end{tabular}

\end{table}
\vspace{-10pt}
\begin{figure}[t]
  \centering
  \includegraphics[width=7.5cm]{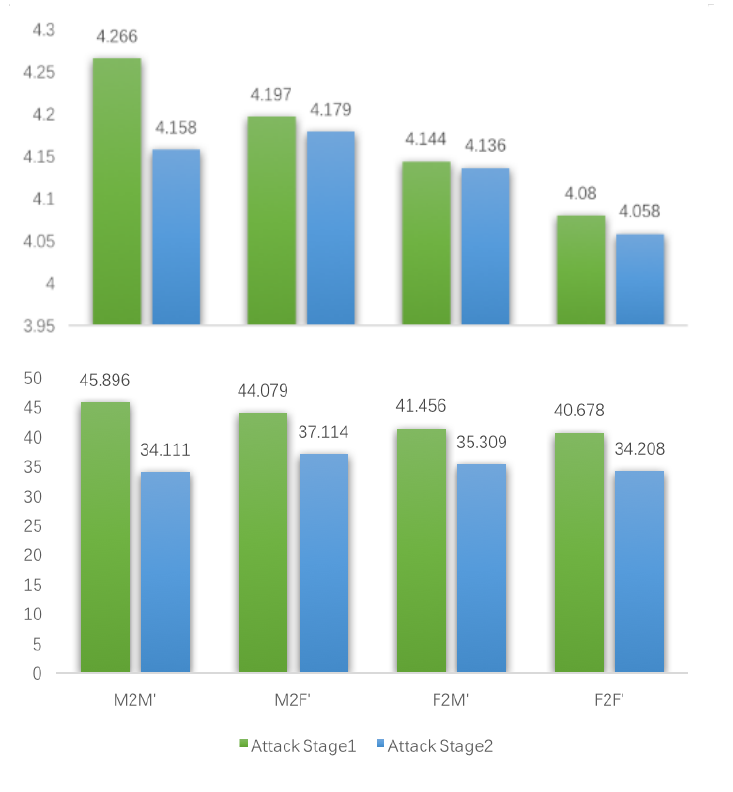}
  \caption{Average PESQ and SNR (dB) comparison of Attack Stage1 and Attack Stage2 on each test mode.}
  \label{fig:snr}
  \vspace{-13pt}
\end{figure}
\subsection{Objective evaluation and subjective listener evaluation}

After conducting the attacks, we want to analyze the adversarial examples from each attack stage. Fig.~\ref{fig:snr} shows objective performance of the generated adversarial examples.
We can observe that the objective performance of the Attack Stage1 adversarial examples is slightly better than Attack Stage2.  
The reason of these results is frequency masking only hide the perturbation in the masking threshold, but does not decrease the energy of the perturbations of the adversarial examples. 
So we also perform subjective test to evaluate the similarity of the adversarial examples and the original wave to find out whether the perturbations generated in Attack Stage2 is inaudible to listeners.

To subjectively evaluate the performance of both attack stages, we conduct ABX preference test.  
In our task, 20 utterances pairs of are chosen randomly from the four test modes as evaluation speech and each pair is judged by 30 participants. The voices for comparison are separately the adversarial examples generated from Attack Stage1 and Attack Stage2.
Participants were asked to make judgement mainly according to
``which one is more similar to the original voice?".

Table~\ref{tab:2} summarizes the ABX test results. We can see that the Attack Stage2 obtains better preference score than the Attack Stage1 ($p$-value$<$0.05). The result indicates that frequency masking make the perturbations more inaudible when generating the adversarial examples, even with larger absolute energy. Some samples of generated adversarial examples can be found on this website\footnote{https://pengchengguo.github.io/inaudible-advex-for-sv}.
\begin{table}[!ht]

  \caption{Preference scores (\%) of Attack Stage1 and Attack Stage2.}
\vspace{-3pt}
  \label{tab:2}

  \centering
\begin{tabular}{*{4}{c}}
\toprule
\multicolumn{3}{c}{Preference (\%)} & \multirow{2}{*}{$p$-value} \\
\cmidrule{0-2} Attack Stage1 & Neural & Attack Stage2  \\
\midrule
11.33 & 20.00 & \textbf{68.67} & 0.0379 \\
\bottomrule
\end{tabular}
\end{table}
\vspace{-10pt}
\subsection{Non-speech targeted attack}
We also use music as the original input to conduct the targeted speaker attack. We match each utterance with a target speaker label and measure the attack success rate.
The result shows in Table~\ref{tab:3}. We first use original music wave with target speaker labels to test the system and get 0.00\% of prediction accuracy. After generating adversarial examples from Attack Stage1 and Attack Stage2, we can achieve 77.0\% and 91.5\% attack success rate, respectively. The experimental result demonstrates the attacking effectiveness of the inaudible adversarial perturbations, even applied to a completely irrelevant waveform.
\begin{table}[!ht]

  \caption{Attack success rate (\%) of Attack Stage1 and Attack Stage2 on non-speech dataset.}
\vspace{-3pt}
  \label{tab:3}

  \centering

  \begin{tabular}{c c c c }

    \toprule
    \multicolumn{1}{c}{\textbf{ }} &
    \multicolumn{1}{c}{\textbf{Before Attack}} &
    \multicolumn{1}{c}{\textbf{Attack Stage1}} &
    \multicolumn{1}{c}{\textbf{Attack Stage2}} \\
    \midrule
    \textit{Acc} & 0.00\% & 77.0\% & \textbf{91.5}\%   \\
    \bottomrule

  \end{tabular}

\end{table}
\vspace{-12pt}
\section{Conclusion}

In this study, we have proposed to targeted attack the speaker recognition system by generating inaudible adversarial perturbations. In particular, the psychoacoustic principle of frequency masking is used for the generation of adversarial examples. We constrict the perturbation under the masking threshold of the original audio, instead of a common $l_p$ distortion measures.
Experiments on Aishell-1 corpus show that our approach yields up to 98.5\% attack success rate to arbitrary gender speaker targets, while retaining indistinguishable attribute to listeners. In subjective listener evaluation, the frequency masking based adversarial perturbations have a 68.67\% preference, which indicates the frequency masking based adversarial perturbations are more inaudible, even with larger absolute energies. Furthermore, the results demonstrate the effectiveness when applying to non-speech data, such as music, to conduct targeted speaker attacks.

In our future work, we will explore more challenging scenarios, both white-box and black-box targeted attacks and the defenses of the adversarial examples. On-the-air targeted attacks~\cite{xie2020real} and defenses also are within our future plan.

\clearpage
\bibliographystyle{IEEEtran}

\bibliography{mybib}

\end{document}